# Ineffectiveness of Formamidine in Suppressing Ultralow Thermal Conductivity in Cubic Hybrid Perovskite $FAPbI_3$

Jiongzhi Zheng[1,2*], Zheng Chang[3,*], Changpeng Lin[4,5], Chongjia Lin[6], Yanguang Zhou[6], Baoling Huang[6,7,8,†], Ruiqiang Guo[9,‡], Geoffroy Hautier[1,§]

[1]*Thayer School of Engineering, Dartmouth College, Hanover, New Hampshire, 03755, USA*
[2]*Energy Technologies Area, Lawrence Berkeley National Laboratory, Berkeley, CA 94720, USA*
[3]*School of Naval Architecture and Maritime, Zhejiang Ocean University, Zhoushan, Zhejiang, 316022, China*
[4]*Theory and Simulation of Materials (THEOS), École Polytechnique Fédérale de Lausanne, CH-1015 Lausanne, Switzerland*
[5]*National Centre for Computational Design and Discovery of Novel Materials (MARVEL), École Polytechnique Fédérale de Lausanne, CH-1015 Lausanne, Switzerland*
[6]*Department of Mechanical and Aerospace Engineering, The Hong Kong University of Science and Technology, Clear Water Bay, Kowloon, Hong Kong*
[7]*HKUST Foshan Research Institute for Smart Manufacturing, Hong Kong University of Science and Technology, Clear Water Bay, Kowloon, Hong Kong, China*
[8]*HKUST Shenzhen-Hong Kong Collaborative Innovation Research Institute, Futian, Shenzhen 518055, China*
[9]*Thermal Science Research Center, Shandong Institute of Advanced Technology, Jinan, Shandong Province, 250103, China*

## Abstract

Understanding lattice dynamics and thermal transport mechanisms in cubic hybrid organic–inorganic perovskites remain challenging due to strong anharmonicity and phase transitions. Here, we investigate the thermal transport behavior in benchmark cubic hybrid perovskite $FAPbI_3$ by coupling first principles-based anharmonic lattice dynamics with a linearized Wigner transport equation. Using the Temperature-Dependent Effective Potential (TDEP) method, we stabilize the negative soft modes, primarily dominated by organic $FA^+$ cations. Our calculations predict an ultra-low thermal conductivity of ~0.63 $\mathrm{Wm^{-1}K^{-1}}$ at 300 K, following a temperature dependence of $T^{-0.740}$. Contrary to common assumptions, we find that the $[PbI_3]^{1-}$ units, rather than $FA^+$ cations, dominate thermal resistance. Furthermore, we demonstrate that anharmonic force constants are highly temperature-sensitive, relying on 0-K force constants significantly underestimates thermal conductivity. Our study not only elucidates the microscopic mechanisms governing thermal transport in $FAPbI_3$ but also provides a robust framework for modeling heat conduction in hybrid organic-inorganic compounds.

[*] These authors contributed equally.
[†] mebhuang@ust.hk
[‡] ruiqiang.guo@iat.cn
[§] Geoffroy.hautier@dartmouth.edu

# INTRODUCTION

Hybrid organic-inorganic halide perovskites have attracted significant attention due to their outstanding semiconducting properties[1–5]. Within the family of metal halide perovskites, formamidinium lead triiodide ($FAPbI_3$) stands out as a promising candidate for efficient and stable perovskite solar cells (PSCs). Compared to the early prototype hybrid perovskite, methylammonium lead triiodide ($MAPbI_3$), formamidinium lead triiodide ($FAPbI_3$) exhibits notable distinctions, particularly its narrower band gap of 1.45 eV and enhanced thermal stability[6–8]. To date, research efforts on formamidinium (FA)-based perovskite materials have primarily centered on stabilizing the $\alpha$-$FAPbI_3$ structure at relatively low temperatures. Many efforts have been dedicated to suppressing the formation of $\delta$ −phase perovskites by promoting their transition to the $\alpha$ −phase, yielding significant success[9–12]. Despite significant advancements in enhancing the photovoltaic energy conversion efficiency and stabilization of α-$FAPbI_3$, concerns remain regarding its vibrational dynamics and thermal transport. Vibrational dynamics impact not only thermal transport[13,14], but also carrier mobilities[15], device performance[16], and thermal instability[17] in perovskites. For instance, the hot-phonon bottleneck effect in lead halide perovskites significantly prolongs the cooling period of charge carriers[18,19]. Furthermore, perovskite materials have recently attracted interest for thermoelectric applications due to their favorable electrical properties[20–22] and ultra-low thermal conductivity[13,14,22–24]. To date, over 1,346 hybrid organic–inorganic perovskites (HOIPs) have been identified and characterized[25]. Thus, exploring the vibrational dynamics and thermal transport in hybrid organic-inorganic perovskites is essential, from both fundamental and practical perspectives.

Recent advancements have significantly enhanced our ability to measure and predict the thermal properties of halide perovskites through experimental and theoretical approaches. Experimentally,

Pisoni *et al.* [24]were the first to report an ultra-low room-temperature thermal conductivity of 0.5 $Wm^{-1}K^{-1}$ in hybrid inorganic-organic halide perovskite $CH_3NH_3PbI_3$. Lee *et al.*[22] conducted experiments that yielded similarly low thermal conductivity of 0.45 $Wm^{-1}K^{-1}$ for all-inorganic halide perovskite $CsPbI_3$ and 0.42 $Wm^{-1}K^{-1}$ and $CsPbBr_3$, respectively. Acharyya *et al.*[13] experimentally observed an ultra-low thermal conductivity ranging from ~0.37 to 0.28 $Wm^{-1}K^{-1}$ for the 2D perovskite $Cs_2PbI_2Cl_2$ across a temperature range of 295 to 523 K. Furthermore, an ultra-low thermal conductivity of approximately 0.20 $Wm^{-1}K^{-1}$ at room temperature, along with a glass-like temperature dependence of thermal conductivity from 2 to 400 K, was experimentally reported in a single crystal of the layered halide perovskite $Cs_3Bi_2I_6Cl_3$[14]. Theoretically, to address the limitations of the harmonic approximation in accurately describing lattice dynamics within highly anharmonic materials, several theoretical methods have been developed, including self-consistent phonon calculations (SCP)[26,27] and the temperature-dependent effective potential (TDEP) approach[28,29]. These theoretical techniques explore the impact of temperature on phonon modes in highly anharmonic perovskite materials, particularly those undergoing phase transitions at finite temperatures. Specifically, the SCP approach was employed to analyze finite-temperature lattice dynamics in oxide perovskites $SrTiO_3$[27] and $BaZrO_3$[30], as well as cubic halide perovskites $CsMBr_3$ (where M = Ca, Cd, and Sn)[31] and $CsPbBr_3$[26]. The TDEP technique was also used to stabilize the complex perovskites such as double perovskite $Cs_2AgBiBr_6$[32], 2D perovskite $Cs_2PbI_2Cl_2$[33] and layered perovskite $Cs_3Bi_2I_6Cl_3$[34].

Given the strong anharmonicity in perovskite materials, it's crucial to go beyond conventional phonon quasiparticle picture considering only three-phonon scattering processes when analyzing thermal transport phenomena. Specifically, four-phonon scattering processes are found to be key to understanding lattice thermal conductivity in these highly anharmonic perovskites, including oxide[30], fluoride[35] and halide variants[33,36]. Moreover, considering the wave-like phonon tunneling channel becomes essential to precisely explain thermal transport in highly anharmonic perovskites[33,34,36–39], particularly when their thermal conductivity approaches the theoretical minimum limit. While only a few theoretical research efforts have focused on thermal transport in hybrid organic-inorganic perovskites, these studies frequently concentrate on the stable phase, often neglecting higher-order scatterings[40,41], or rely on empirical potential molecular dynamics simulations[42–44]. Hence, accurately predicting lattice dynamics and gaining a microscopic understanding of thermal transport in the high-temperature phase of photoactive cubic hybrid organic-inorganic perovskites remain in their infancy, underscoring the urgent need for further research.

In this work, we thoroughly investigate the temperature-dependent lattice dynamics and the microscopic mechanisms of thermal transport in the cubic hybrid organic-inorganic perovskite $FAPbI_3$. Cubic hybrid crystalline $FAPbI_3$ is a promising candidate for efficient and relatively stable perovskite solar cells and is therefore selected as the benchmark system in the current work. We employ a state-of-the-art first-principles framework that integrates the Temperature-Dependent Effective Potential (TDEP) approach with the Wigner transport formula to assess the thermal transport properties of cubic $FAPbI_3$. This framework incorporates both three-phonon (3ph) and four-phonon (4ph) scatterings within the diagonal and non-diagonal terms of the heat flux operators, thereby providing a robust depiction of thermal transport phenomena in cubic $FAPbI_3$. In zero-K phonon calculations, we observe dynamical instability associated with the $FA^+$ cations in cubic $FAPbI_3$. Further, we demonstrate the impact of 4ph scatterings on particle-like phonon propagation and wave-like tunnelling of phonons and predict an ultra-low thermal conductivity for cubic $FAPbI_3$. Through meticulous investigation of the mesoscopic mechanisms of thermal transport, we pinpoint the origin of ultra-low thermal conductivity to the $[PbI_3]^-$ units, rather than the $FA^+$ cations, in cubic $FAPbI_3$. Despite its complex structure and strong anharmonicity, cubic $FAPbI_3$ primarily exhibits thermal conductivity through the particle-like phonon propagation channel. Finally, we analyze the sensitivity of the anharmonic force constants to temperature, underscoring the importance of extracting all force constants at finite temperatures in hybrid perovskites. These results provide a comprehensive understanding of heat conduction in the cubic $FAPbI_3$ structure, thus advancing the knowledge of thermal transport in hybrid organic-inorganic compounds.

## RESULTS

### Crystal Structure and temperature-dependent Phonon Dispersions and DOS

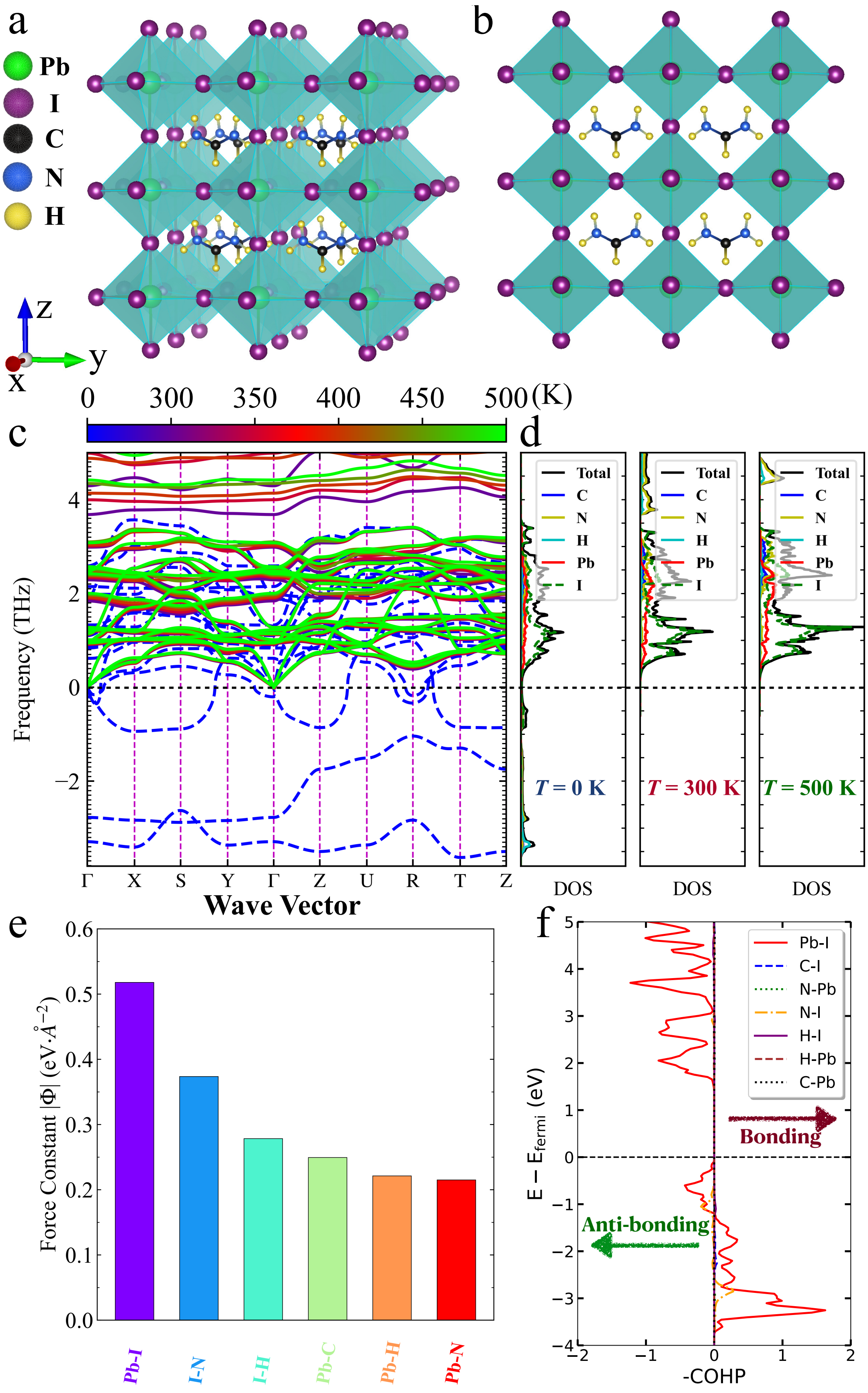

**Fig. 1. Crystal structure, phonon dispersions and DOS, Harmonic IFCs and COHP. a** Three-dimensional (3D) schematic representation of the crystal structure of cubic hybrid organic-inorganic perovskite $FAPbI_3$, characterized by corner-sharing $[PbI_3]^-$ octahedra units and an organic cation $FA^+$. In the diagram, Pb, I, C, N, and H atoms are color-coded as green, purple, black, blue, and yellow, respectively. **b** Perspective view of the crystal structure of the yz plane. **c** Comparison of finite-temperature phonon dispersions calculated from 300 to 500 K with those obtained from the harmonic approximation treatment at 0 K. **d** Atom-decomposed partial and total phonon density of states (DOS) calculated at 0 K, 300 K, and 500 K, respectively. Here, the full phonon dispersions and DOS, please refer to **Supplementary Figure 1** in Supplementary Information (SI). **e.** The norm of harmonic IFCs for nearest neighbor atomic pairs calculated at 300 K. **f** The crystal orbital Hamilton population (COHP) of atomic pairs in cubic crystalline $FAPbI_3$. Here, the negative and positive values indicate the anti-bonding and bonding states, respectively.

We start by analyzing the crystal structure and both the harmonic and anharmonic lattice dynamics of the cubic hybrid organic-inorganic perovskite $FAPbI_3$, as depicted in Figs. 1(a-d). In the cubic framework of crystalline $FAPbI_3$[45], $Pb^{2+}$ ions occupy the interstitial sites within the tetrahedrally coordinated sublattice formed by iodine (I) atoms, leading to the formation of $[PbI_3]^{1-}$ octahedra units. The planar organic $FA^+$ ($[H_2N\text{-}CH\text{-}NH_2]^+$) cation resides at the center of the cube, surrounded by four corner-sharing $[PbI_3]^{1-}$ octahedra units, with the alignment of the C-H bond along <100>[45] [see Figs. 1(a-b)].

Using the harmonic approximation treatment[46], we calculate the vibrational properties of cubic crystalline $FAPbI_3$ at zero Kelvin. In Figures.1(c-d), the prominent feature is the presence of several phonon branches exhibiting imaginary frequencies, suggesting the dynamical instability of cubic $FAPbI_3$ at low temperatures, which aligns with experimental findings[45]. The unstable modes primarily originate from the $FA^+$ cations, specifically the H atoms, as evidenced by the atomic decomposed partial DOS and the projected atomic participation ratio [See Figs. 1(c-d) and **Supplementary Figure 2(a)** in the (SI)]. This phenomenon can be attributed to the weak bonding between the $FA^+$ cations and the $[PbI_3]^-$ units [see Fig. 1(e)], leading to the random orientation of $FA^+$ cations, as depicted in **Supplementary Figure 3(a-c)** in the SI. This observation contrasts with inorganic halide perovskites[32,36,47], where the unstable modes are predominantly driven by the tilting of the tetrahedrally coordinated anions, such as $[BiBr_6]$ and $[AgBr_6]$ units. After careful investigation, we find that the phonon modes with small imaginary frequencies at the **R** point are predominantly associated with $[PbI_3]^-$ units [see Figs. 1(c) and **Supplementary Figure 2(b)** in the SI], corresponding to induced phase-transition modes in oxide and halide perovskites[26,27]. To gain deeper insight into $PbI_6$ octahedra-induced soft modes, we also calculate the crystal orbital Hamiltonian population (COHP) for cubic crystalline $FAPbI_3$, illustrated in Fig. 1(f). Near the Fermi energy level, we observe the presence of anti-bonding states contributed by the Pb-I bonding, which typically results in strong anharmonicity, weak bonding and soft phonon modes[48,49]. As a result, the strong random orientation of $FA^+$ cations induces tilting of the $PbI_6$ octahedra, causing the deformation of cubic $FAPbI_3$ into an unfavorable phase (δ phase) as temperature decreases.

To account for temperature effect (lattice anharmonicity) on phonons in cubic crystalline $FAPbI_3$, the temperature-dependent effective potential approach[28,29] was employed to anharmonically renormalize phonon energies at finite temperatures. From Figs. 1(c-d), we observe that all anharmonically renormalized phonons exhibit stabilization above 300 K, consistent with the experimental phase transition temperature of 300 K[45]. While light elements, namely H and N atoms, govern the unstable modes with large negative frequencies at zero Kelvin, heavy elements such as I and Pb atoms dominate the low-frequency phonon modes ($<\sim$2.5 THz) at finite temperatures [see Fig. 1(c-d)]. This observation further underscores the pivotal role of $FA^+$ cations in inducing the tilting of $PbI_6$ octahedra, thereby facilitating the phase transition between the α and δ phases of $FAPbI_3$. Furthermore, with increasing temperature (>300 K), the I-dominated phonon modes, i.e., the low-frequency optical modes ($\leq$ 2.5 THz), experience a gradual progressive stiffening. This stiffening is observed to be relatively weak, as depicted in Figs. 1(c-d), in striking contrast to the behavior observed in the double perovskite $Cs_2AgBiBr_6$[32,36]. However, the H-dominated high-frequency modes (> 3.5 THz) exhibit significant phonon stiffening, suggesting the presence of strong four-phonon processes[27,50], as shown in Figs. 1(c-d). The strong four-phonon scatterings of high-frequency phonon modes can be attributed to the large mean-square atomic displacements (MSD) of $FA^+$ cations, which contributes to the rattling-like motion of the H and N atoms [see **Supplementary Figure 4(a-e)** in the SI]. It is worth noting that the rattling-like modes from $FA^+$ cations impact thermal transport differently compared to those from the heavy metallic A site of other perovskites[32,36,51,52], a topic that will be discussed later.

## Potential energy surfaces and animations

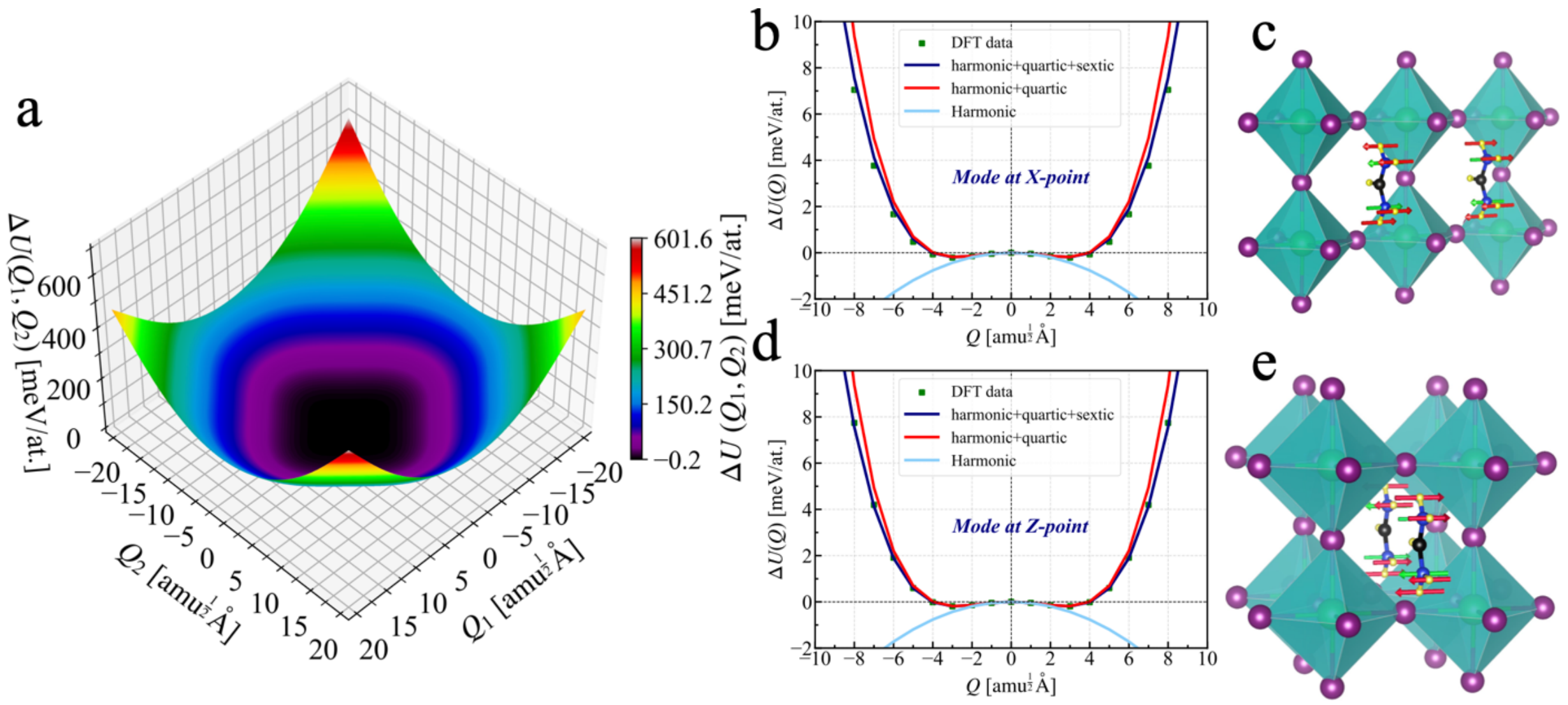

**Fig. 2. Potential energy surfaces and lattice modal animations. a** Calculated two-dimensional (2D) potential energy surface of cubic $FAPbI_3$ associated with normal mode coordinates $Q_1$ and $Q_2$. The lowest soft modes at **Γ** − and **X-** points were utilized to generate the potential energy surfaces. **b** Calculated one-dimensional (1D) potential energy surface of lowest mode at **Γ**-point as a function of normal mode coordinate $Q_1$. **c** The lattice vibrational animation associated with lowest modes at **Γ**-point. **d** Calculated one-dimensional (1D) potential energy surface of lowest mode at **X-**point as a function of normal mode coordinate $Q_2$. **e** The lattice vibrational animation associated with lowest modes at **X-**point.

To gain an intuitive insight into lattice anharmonicity and instability in the cubic $FAPbI_3$, we calculate the potential energy surfaces (PES)[53] for the large imaginary modes at the $\Gamma -$ and $\mathrm{X} -$ points, as illustrated in Figs. 2(a-e). Both potential energy surfaces (PESs) exhibit a deep double well with a relatively flat bottom, suggesting strong anharmonicity for the unstable phonon modes[54]. In particular, the energy minima are situated outside of the zero-tilt amplitude ($Q_1$=$Q_2$=0) for both soft modes at the Γ and X points, respectively, indicating dynamical instability of cubic $FAPbI_3$ at zero K, as shown in Fig. 2(a). Especially, the harmonic approximation fails to capture the U-shaped double-well PESs, while a fourth-order polynomial (four-phonon scattering processes) can be used to accurately reproduce the actual PESs, as illustrated in Figs. 2(b) and (d). This phenomenon was also observed in other crystals such as $BaZrO_3$[30] and the double perovskite $Cs_2AgBiBr_6$[36], as well as in clathrate $Ba_8Ga_{16}Ge_{30}$[50], highlighting the importance of higher-order phonon scattering processes in determining effective phonon energies. Clearly, both double-well U-shaped potential energy surfaces (PESs) are exclusively associated with the sublattice rotation of $FA^+$ cations, which aligns with the varying orientation of $FA^+$ cations in different phases of crystalline $FAPbI_3$[55].

## Lattice thermal conductivity and Phonon scattering properties

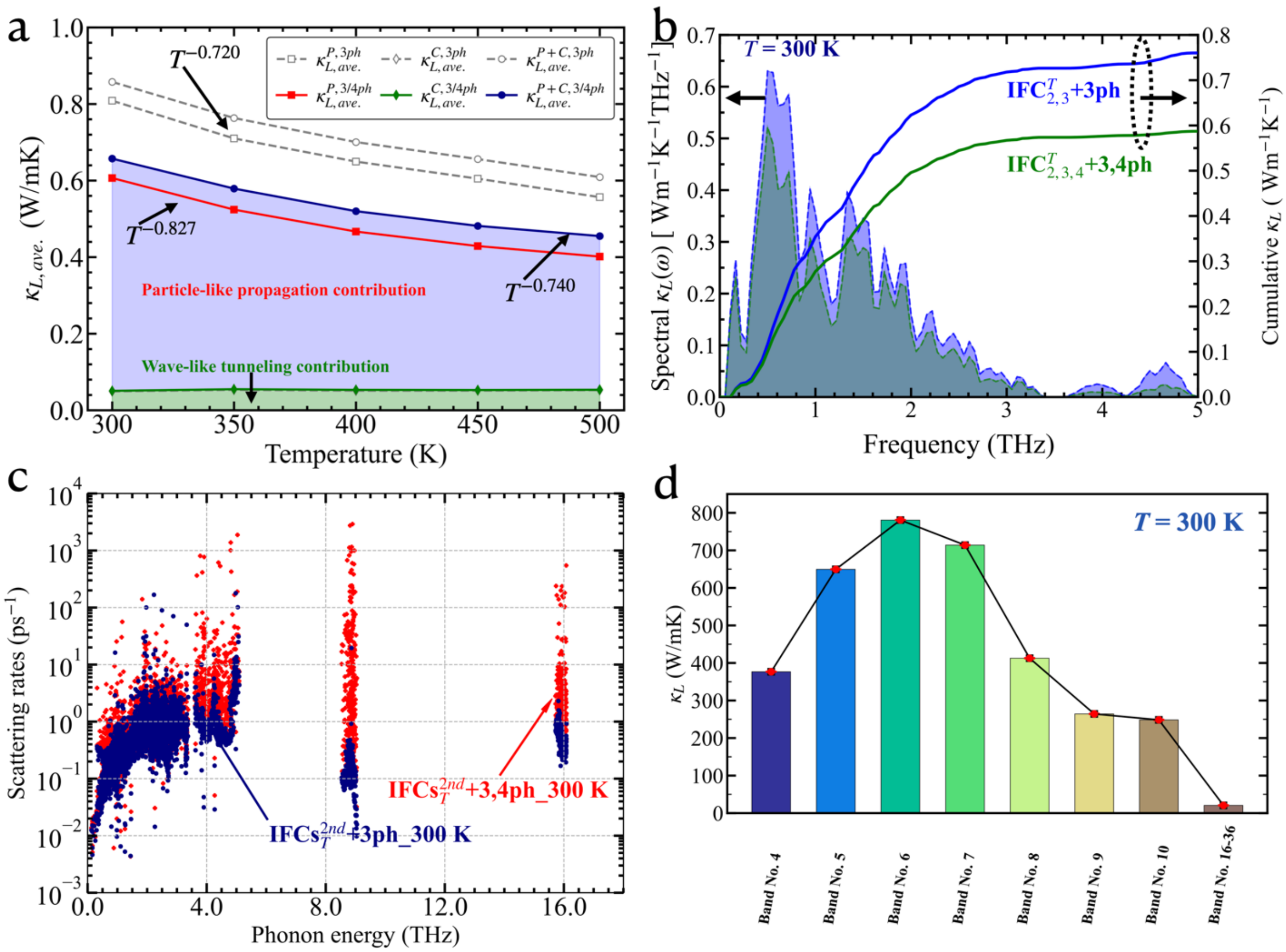

**Fig.3. Lattice thermal conductivity, phonon scattering rates and phase spaces.** **a** Calculated temperature-dependent averaged lattice thermal conductivity, including contributions from population and coherence conductivities, accounts for only 3ph both 3ph and 4ph scattering processes. The light blue shaded area indicates the particle-like phonon propagation contributions, and the light green shaded area represents the wave-like phonon tunnelling contributions. **b** Calculated averaged spectral and cumulative lattice thermal conductivity from particle-like phonon propagation channel considering only 3ph, and both 3ph and 4ph scattering processes at 300 K. **c** Calculated 3ph and both 3ph and 4ph scattering rates at 300 K, respectively. Here, the harmonic ($2^{nd}$-) and anharmonic force constants ($3^{rd}$- and $4^{th}$-order) are extracted at finite temperature (300 K). **d** Calculated population thermal conductivity at 300 K using phonon dispersions, with selective exclusion of specific phonon bands. Herein, elements of the scattering matrix involving the omitted modes are set to zero, isolating their effects on thermal transport.

With finite-temperature IFCs at hand, we proceed to calculate the thermal conductivity $\kappa_{\mathrm{L}}$ in cubic $FAPbI_3$ using the advanced thermal transport model, the Wigner transport formula, which incorporates population $\kappa_{\mathrm{L}}^{\mathrm{P}}$ and coherence contributions $\kappa_{\mathrm{L}}^{\mathrm{C}}$ [38,39]. It is noteworthy that all IFCs, including both harmonic and anharmonic terms, are extracted at finite temperatures and utilized to evaluate the lattice $\kappa_{\mathrm{L}}$ for cubic crystalline $FAPbI_3$. When considering only three-phonon (3ph) scattering processes, we predict an ultra-low $\kappa_{\mathrm{L}}^{\mathrm{P}}$ of 0.75 $\mathrm{Wm^{-1}K^{-1}}$ at 300 K and 0.4 $\mathrm{Wm^{-1}K^{-1}}$ for cubic $FAPbI_3$ at 500 K, as illustrated in Fig. 3(a). As previously discussed, higher-order anharmonicity plays a crucial role in determining effective phonon energies. Consequently, it should also exert a substantial influence on phonon scattering rates. Further incorporating the effect of four-phonon (4ph) scatterings, the predicted $\kappa_{\mathrm{L}}^{\mathrm{P}}$ decreases to 0.64 at 300 K and 0.3 $\mathrm{Wm^{-1}K^{-1}}$ at 500 K, representing a 25% and 34% reduction, respectively [see Fig. 3(a)]. Interestingly, the reduction in thermal conductivity due to 4ph scatterings in cubic $FAPbI_3$ is significantly lower than that observed in double perovskite $Cs_2AgBiBr_6$[36]. This difference can be attributed to the absence of phonon modes from the A site with heavy metallic elements in cubic $FAPbI_3$ [see Fig. 1(d)], which are typically involved in flattening phonon branches and thus result in strong four-phonon scatterings[36].

To gain a deeper insight into population thermal conductivity $\kappa_{\mathrm{L}}^{\mathrm{P}}$ in cubic $FAPbI_3$, we analyze both the spectral and cumulative $\kappa_{\mathrm{L}}^{\mathrm{P}}$, as depicted in Fig. 3(b). It is evident that phonons with frequencies less than 2 THz primarily dominate the populations' conductivity $\kappa_{\mathrm{L}}^{\mathrm{P}}$ in cubic $FAPbI_3$. From Fig. 1(d), we observe that the phonons with frequencies less than 2 THz are dominated by I and Pb atoms, indicating the critical role of $[PbI_3]^-$ units on contributing thermal conductivity in cubic crystalline $FAPbI_3$. Notably, phonons with frequencies below 2 THz play a crucial role in suppressing $\kappa_{\mathrm{L}}^{\mathrm{P}}$ through 4ph scatterings. More specifically, multiple dips in spectral $\kappa_{\mathrm{L}}^{\mathrm{P}}(\omega)$ around phonons at 1 and 2 THz suggest strong phonon scattering rates[32,56], which significantly contribute to the ultra-low thermal conductivity observed in cubic $FAPbI_3$. We also identify several peaks (long tails) in phonon scattering rates, attributable to nearly flattened phonon bands around 1 and 2 THz[57–59], as depicted in Figs. 1(c) and 3(c). In Figure 3(c), four-phonon scatterings significantly dominate over three-phonon scatterings for phonons with frequencies larger than 3.5 THz, due to the flattening phonon bands [see Fig. 1(c)]. Despite the strong 4ph scattering, phonons with frequencies higher than 3.5 THz, dominated by $FA^+$ cations [see Figs. 1(d) and 3(c)], do not significantly impact the suppression of $\kappa_L^P$ [see Fig. 3(b)].

To assess the influence of specific modes on thermal transport suppression in cubic $FAPbI_3$, we computed the population conductivity by isolating the corresponding mode, namely, by setting the elements of the scattering matrix that involve this mode to zero, as depicted in Fig. 3(d). The similar analysis was also conducted for lead-free double perovskite $Cs_2AgBiBr_6$[36]. Phonon modes within the frequency range of 0.8 to 2 THz, corresponding to bands 4-10 and dominated by the I and Pb atoms [see Fig. 1(d)], have a substantial effect on thermal transport. Conversely, modes with frequencies exceeding 3.5 THz, associated with bands No. 16-32 and C, N and H atoms, exert a considerably less impact on thermal transport within cubic $FAPbI_3$. Notably, exclusion of band No. 6 (~1 THz) results in a dramatic increase in $\kappa_L^P$ from 0.75 $Wm^{-1}K^{-1}$ to 780 $Wm^{-1}K^{-1}$. In contrast, excluding bands No. 16-36 (> ~3.5 THz) leads to only a modest increase in $\kappa_L^P$ from 0.75 $Wm^{-1}K^{-1}$ to only 20 $Wm^{-1}K^{-1}$. This finding underscores that phonon modes with frequencies below ~ 2 THz exhibit strong anharmonicity and are pivotal in suppressing the ultra-low thermal conductivity in cubic $FAPbI_3$. Importantly, our findings highlight that the $[PbI_3]^-$ units play the dominant role in suppressing thermal transport in cubic crystalline $FAPbI_3$, compared to the $FA^+$ functional group, which has a minimal effect on thermal transport. Considering the strong anharmonicity in cubic $FAPbI_3$, we further calculate the coherence contributions $\kappa_L^C$ from wave-like phonon tunnelling channel[38,39], as depicted in Fig. 3(a). Although strong anharmonicity is observed in cubic $FAPbI_3$, the contribution of $\kappa_L^C$ is found to be minor, accounting for only 10% and 15% of the total $\kappa_L$ at 300 and 500 K, respectively. This result not only emphasizes the dominant role of populations' contribution to the total $\kappa_{\mathrm{L}}$ but also confirms the limited impact of strong anharmonicity from $FA^{+}$ cations on enhancing $\kappa_{\mathrm{L}}^{\mathrm{C}}$ in cubic $FAPbI_3$.

We next examine the temperature dependence of lattice thermal conductivity $\kappa_{\mathrm{L}}$, as illustrated in Fig. 3(a). Within the framework of three-phonon scatterings, the predicted $\kappa_{\mathrm{L}}^{\mathrm{P}}$ exhibits a weak temperature dependence of $\sim T^{-0.720}$, deviating from the conventional temperature dependence of $\sim T^{-1}$ [60]. This variation can be attributed to anharmonic phonon renormalization, which reduces phonon scatterings, a phenomenon also observed in compounds such as $BaZrO_3$[30] and $Cs_2AgBiBr_6$[36]. With the inclusion of the effect of 4ph scatterings, the temperature dependence of $\kappa_{\mathrm{L}}^{\mathrm{P}}$ becomes stronger, following $\sim T^{-0.827}$, due to the stronger temperature dependence of 4ph scattering as compared to 3ph scattering[30,36,61]. Moreover, the contribution to the $\kappa_{\mathrm{L}}$ from wave-like phonon tunnelling channel is minor in cubic $FAPbI_3$, resulting in little change in its temperature dependence. This is in sharp contrast to previous observation that considering $\kappa_{\mathrm{L}}^{\mathrm{C}}$ in highly anharmonic compounds often significantly modifies the temperature dependence of total thermal conductivity, potentially leading to temperature independence at high temperatures[62], or even a positive temperature dependence[63]. This finding indicates the good crystalline nature of phonon transport in cubic $FAPbI_3$, despite containing a complex organic functional group, namely $FA^{+}$ cations.

## Projected phonon dispersions and electronic band structure.

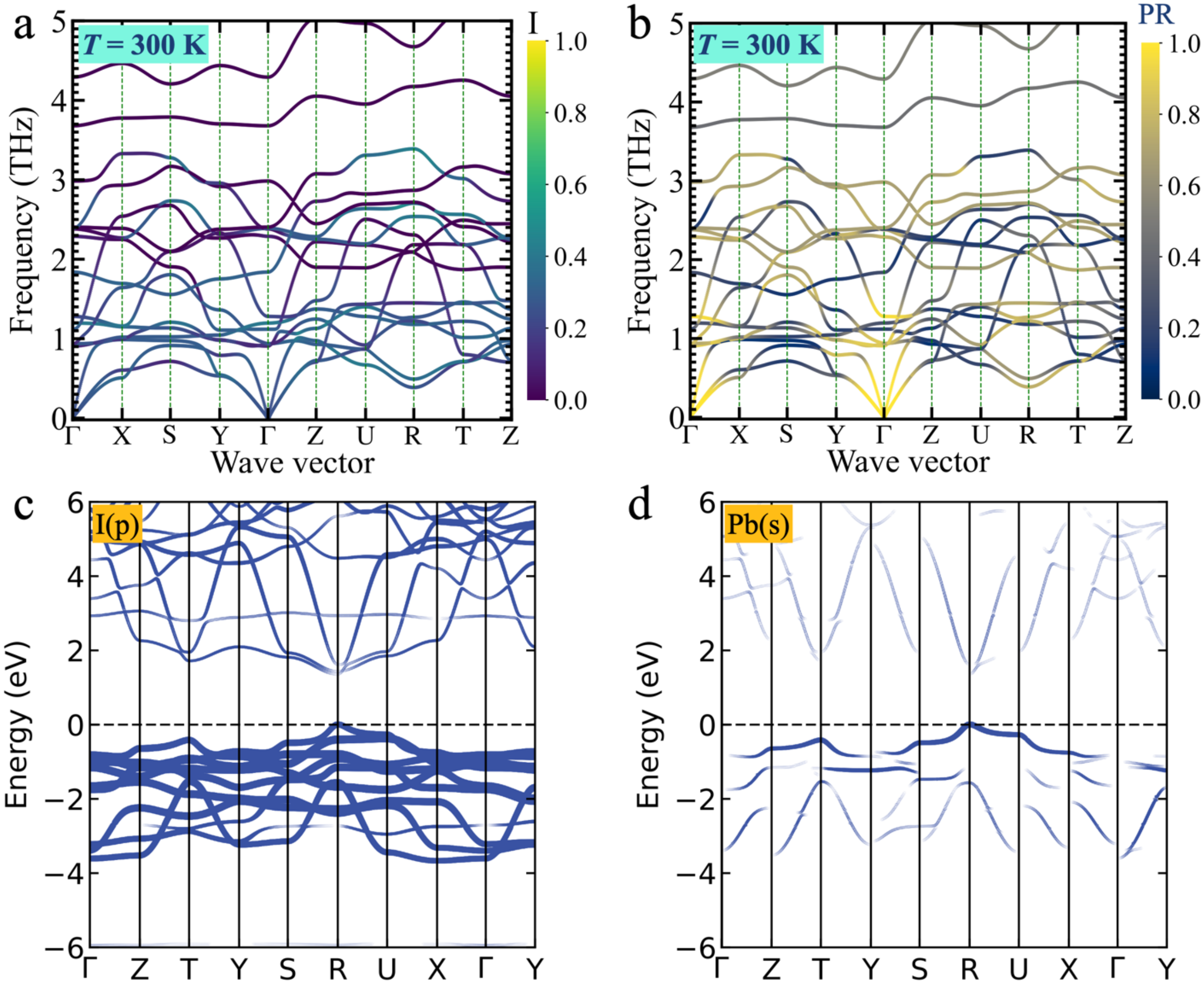

**Fig. 4. Projected phonon dispersions and electronic band structures. a.** The color-coded atomic participation ratio (APR) of cubic crystalline $FAPbI_3$, projected onto the phonon dispersions along the high-symmetry paths at 300 K. The fraction displayed in the color bar indicates the atomic participation ratio of iodine (I) atom on specific phonons. **b**. The color-coded participation ratio of cubic crystalline $FAPbI_3$, projected onto the phonon bands at 300 K. The fraction displayed in the color bar represents the participation ratio of all atoms in primitive cell on specific phonons. **c**. The orbital projected electronic band structure along high-symmetry paths, illustrating the contributions of the I(5p) states. **d**. The same as (**c**), but for the Pb (6s) states. Here, the full electronic band structure and DOS, please refer to **Supplementary Figure 5** in the SI.

To further elucidate the origin of the ultra-low thermal conductivity $\kappa_L$ in the cubic hybrid organic-inorganic perovskite $FAPbI_3$, we analyzed both the atomic and total atomic participation ratios projected onto the phonon bands, as detailed in Figs. 4(a) and (b). As demonstrated in Fig. 4(a), phonons with frequencies below 2 THz are predominantly influenced by iodine (I) atoms, which primarily interact with the main heat carriers, namely acoustic phonons. Additionally, the low participation ratios observed in the phonon bands, as shown in Fig. 4(b), correspond to the iodine-dominated bands. After careful investigation, we find that these low-participation ratio phonon modes can be traced back to the rotational motion of octahedral unit $[PbI_3]^-$, which is also observed in other perovskites, such as octahedral unit $[AgBr_3]^-$ and $[BiBr_3]^-$ in double perovskite $Cs_2AgBiBr_6$[36], indicating strong anharmonicity.[see the animation in **Supplementary Figure 6 in Supplemental Information**]. Therefore, phonon modes with frequencies below 2 THz are associated with strong scatterings, which contribute to the ultra-low thermal conductivity observed in cubic FAPbI3. This phenomenon can be attributed to the significant anharmonicity of the iodine-dominated phonon modes. In contrast, the mechanism behind the ultra-low $\kappa_L$ in the low-temperature phase of other organic-inorganic perovskites like $MAPbI_3$ primarily involves the organic cations ($MA^+$), which serves as the main source of thermal transport suppression[64]. Specifically, Kovalsky et al.[64] identified that resonant phonon frequencies between ~0.45 to 0.90 THz, associated with the hindered rotational degree of freedom of the organic ion, were crucial in suppressing thermal transport in the low-temperature complex phase of $MAPbI_3$. However, in cubic $FAPbI_3$, the $FA^+$ cations mainly contribute to the phonons with frequencies larger than 3.5 THz due to their light masses [see Fig. 1(c-d)], resulting in limited scattering interactions with the primary heat carriers. Therefore, the organic cations, i.e., $FA^+$ cations, play a minor role in suppressing thermal transport in cubic crystalline $FAPbI_3$.

To gain deeper insights into the origin of ultra-low $\kappa_L$ from electronic states in cubic $FAPbI_3$, we projected the atomic electronic orbitals onto the electronic band structures, as depicted in Figs. 4(c-d). The electronic states near the Fermi energy level are primarily contributed by the I(p) and Pb(s) orbitals, indicating active states involved in chemical bonding [see Figs. 4(c-d)].

Furthermore, from the COHP plot in Fig. 1(e), we observe that the I(p) orbitals, in conjunction with Pb(s) orbitals, contribute to forming the antibonding states, typically resulting in the strong anharmonicity and ultra-low $\kappa_L$ in compounds[48,49]. To further support our conclusion, we calculated the Integrated Projected Crystal Orbital Hamilton Population (IpCOHP) for both the $FA^+$ organic functional group and the $PbI_3^-$ units. Since larger absolute IpCOHP values indicate stronger bonding, the results clearly show that bonding involving the $FA^+$ cation is significantly stronger than that within the $[PbI_3]^-$ anionic framework (see **Supplementary Table 1 and Figure 7** in the supplementary information). By combining this with evidence from atomic participation ratios [see Figure 4a], which identify the atoms that contribute most significantly to lattice vibrations, we conclude that the inherently weaker bonding in the $[PbI_3]^-$ framework is the primary reason for the ultra-low lattice thermal conductivity of $FAPbI_3$, outweighing the influence of the $FA^+$ cation. This observation underscores the significant role of the tilting of $FA^+$ in suppressing ultra-low thermal conductivity $\kappa_L$ in cubic hybrid organic-inorganic perovskite $FAPbI_3$.

To this end, and to generalize the microscopic mechanisms of thermal transport in perovskites, we compare cubic crystalline $FAPbI_3$ with other perovskite systems ranging from simple to complex, and from organic to hybrid, as summarized in Table 1. Our previous study on the double perovskite $Cs_2AgBiBr_6$ rigorously demonstrated that although Cs-rattling behavior appears in the low-frequency regime, the lower-frequency modes originating from the octahedra $[BX_6]^-$ units play a more significant role in suppressing thermal transport[36]. Based on phonon dispersion and density of states analyses for $Cs_2SnI_6$[37] and $Cs_2PbI_2Cl_2$[13,33], similar behavior was observed, consistent with that of $Cs_2AgBiBr_6$[36]. We therefore conclude that in other double and layered perovskites, the $[BX_6]^-$ octahedra units also contribute more substantially to thermal transport suppression than Cs rattling. For simple perovskites such as $CsBBr_3$ (B = Ca, Cd, Sn), the $[BBr_3]^-$ octahedra contribute to soft modes occupying the low-frequency regime, which more easily interact with heat carriers (i.e., acoustic phonons) compared to Cs vibrations[30]. Thus, the $[BBr_3]^-$ octahedra units are critical in suppressing thermal transport. In general, the $[BX_6]^-$ (or $[BX_3]^-$) units consistently play a key role in reducing thermal conductivity, even though heavy Cs atoms contribute to low-frequency rattling modes. In systems lacking significant A-site phonon modes in the low-frequency regime, such as cubic $FAPbI_3$, the octahedra units become the sole contributors to thermal transport suppression, as demonstrated by the phonon mode elimination technique [see Fig. 3(d)]. Therefore, we conclude that in both cubic inorganic and hybrid organic–inorganic perovskites, the $[BX_6]^-$ units serve as the primary structural feature responsible for limiting thermal conductivity. We note that classical MD simulations on cubic $MAPbI_3$[43] report an opposite trend compared to our findings for cubic $FAPbI_3$. We attribute this discrepancy to methodological differences between the studies.

**Table 1.** Comparison of the dominant phonon scattering mechanisms across a range of perovskite compounds, from simple to complex structures.

| Compound | Main Origin of Scattering | Secondary Origin of Scattering | Theoretical Support / Evidence | Method | No. of Atoms |
|---|---|---|---|---|---|
| **$CsBBr_3$ (B = Ca, Cd, Sn)**[31] | $BBr_6$ | Cs rattling behavior | DFT + WTE | DFT + WTE | 5 |
| **$Cs_2PbI_2Cl_2$**[13] | $PbI_2Cl_4$ | Cs rattling behavior | Yes / participation ratio | DFT | 7 |
| **$Cs_2SnI_6$**[37] | $SnI_6$ | Cs rattling behavior | Yes / participation ratio | DFT + WTE | 9 |
| **$Cs_2AgBiBr_6$**[36] | $AgBr_6$, $BiBr_6$ | Cs rattling behavior | Yes / participation ratio + neglecting mode technique | DFT + WTE | 10 |
| **$FAPbI_3$ [this work]** | $PbI_6$ | None | Yes / participation ratio + neglecting mode technique | DFT + WTE | 12 |
| **$MAPbI_3$**[43] | $MA^+$ | $[PbI_3]^-$ | Yes | CMD + GK | 12 |

## Phonon lifetimes and Two-dimensional modal coherence conductivity

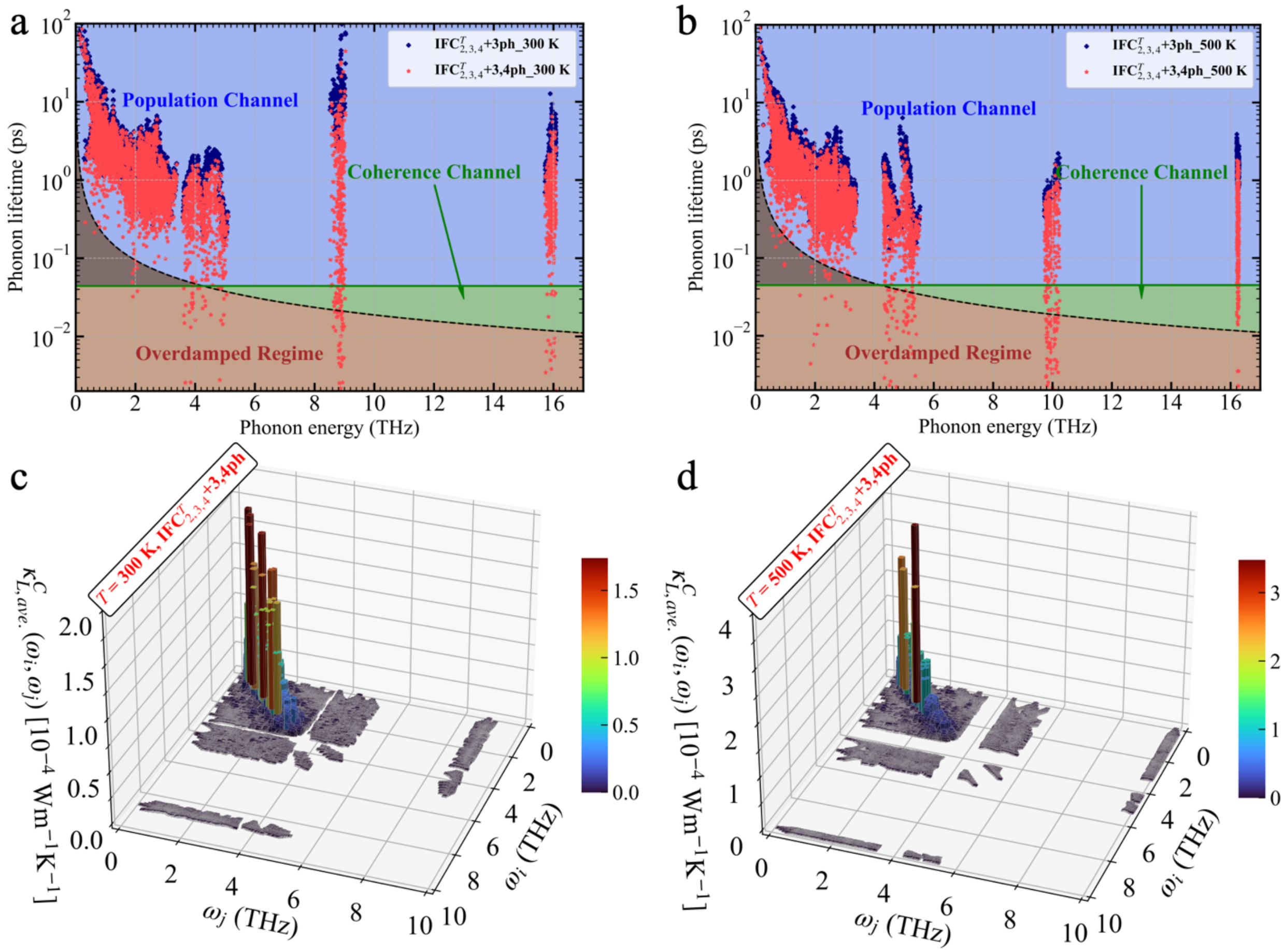

**Fig. 5. Calculated phonon lifetimes and Two-dimensional (2D) modal coherence thermal conductivity. a** Calculated phonon lifetimes considering only 3ph or both 3ph and 4ph scattering rates at 300 K. The solid green horizontal line represents the Wigner limit in time[39]. Phonons with lifetimes above this limit primarily contribute to the population conductivity, while those below it contribute to coherence conductivity. The dash black line indicates the Ioffe-Regel limit in time[65]. **b** The same as (a), but at 500 K. **c** Calculated 2D modal coherence conductivity from wave-like phonon tunnelling considering both 3ph and 4ph scatterings at 300 K. **d** The same as (c), but at 500 K.

Despite the strong anharmonicity in cubic $FAPbI_3$, particularly notable in the strong 4ph scattering rates from $FA^+$ cations [see Figs. 3(c)], the coherence contributions to the total $\kappa_L$ are minor [see Fig. 3(a)]. To elucidate the minor contributions from wave-like phonon tunneling to the thermal transport in cubic $FAPbI_3$, we employed both the Wigner[39] and Ioffe-Regel limit[65] in time to characterize the phonons lifetimes. Phonons with lifetimes exceeding the Ioffe–Regel limit maintain well-defined Lorentzian-shaped phonon spectral functions, which is a key assumption underlying the Wigner transport equation[38,39]. As evidenced in Figs. 5(a) and (b), the majority of phonons exceed the Ioffe-Regel limit, affirming the validity of Wigner transport formula in assessing thermal transport in cubic $FAPbI_3$. Moreover, at temperatures of 300 and 500 K, lifetimes of most of phonons also surpass the Wigner limit, highlighting the dominant role of population contributions in thermal transport in cubic $FAPbI_3$. This phenomenon is due to the large inter-band spacings, owing to the light masses of $FA^+$ cations, which lead to high phonon frequencies. Interestingly, the low-temperature phase of $MAPbI_3$ presents a contrasting scenario wherein coherence contributions dominate the total $\kappa_{\mathrm{L}}$, reflecting its complex crystalline structure[43]. Additionally, although cubic crystalline $FAPbI_3$ is structurally complex (with 12 atoms per unit cell), it exhibits only minor coherence contributions, unlike what has been reported for cubic $MAPbI_3$[66]. This discrepancy may arise from symmetry breaking in the simulation of $MAPbI_3$, where Yang et al. [66] modeled it using larger supercells with lower symmetry space groups (P4/mnc, I4/m, and P1). In contrast, cubic $FAPbI_3$ retains its high-symmetry $Pm\overline{3}m$ structure. It is worth noting that symmetry breaking generally leads to structural complexity and denser phonon mode spacing, thereby enhancing coherence contributions. A similar trend is observed in the non-cubic perovskite $CsPbBr_3$, where the low-symmetry structure contains 20 atoms[39]. Similarly, the complex low-temperature phase of $CsPbBr_3$ exhibits dominant coherence contributions to the total thermal conductivity[39,67], whereas its cubic phase shows that phonon propagation is the primary contributor[67].To further investigate the role of coherence contributions in the thermal transport of cubic $FAPbI_3$, we compute the two-dimensional modal $\kappa_{\mathrm{L}}^{\mathrm{C}}$ at 300 and 500 K, respectively, as illustrated in Figs. 5(c-d). The coherence contributions are primarily driven by quasi-degenerate phonons with frequencies less than ~2 THz, predominantly influenced by iodine (I) atoms [see Fig. 1(c)]. In contrast, phonons with frequencies exceeding 2 THz contribute minimally to the lattice thermal conductivity, due to the wide inter-band spacings, as illustrated in Figs. 5(c-d). This observation suggests that despite exhibiting large four-phonon scatterings, $FA^+$ cations in cubic $FAPbI_3$ minimally impact heat conduction through the wave-like phonon tunnelling channel [see Fig. 3(c)]. As previously discussed, the presence of $FA^+$ cations in cubic $FAPbI_3$ neither significantly affects the phonon population contributions, nor substantially enhances the coherence contributions to the total $\kappa_{\mathrm{L}}$.

## Thermal transport properties calculated using zero-K anharmonic IFCs

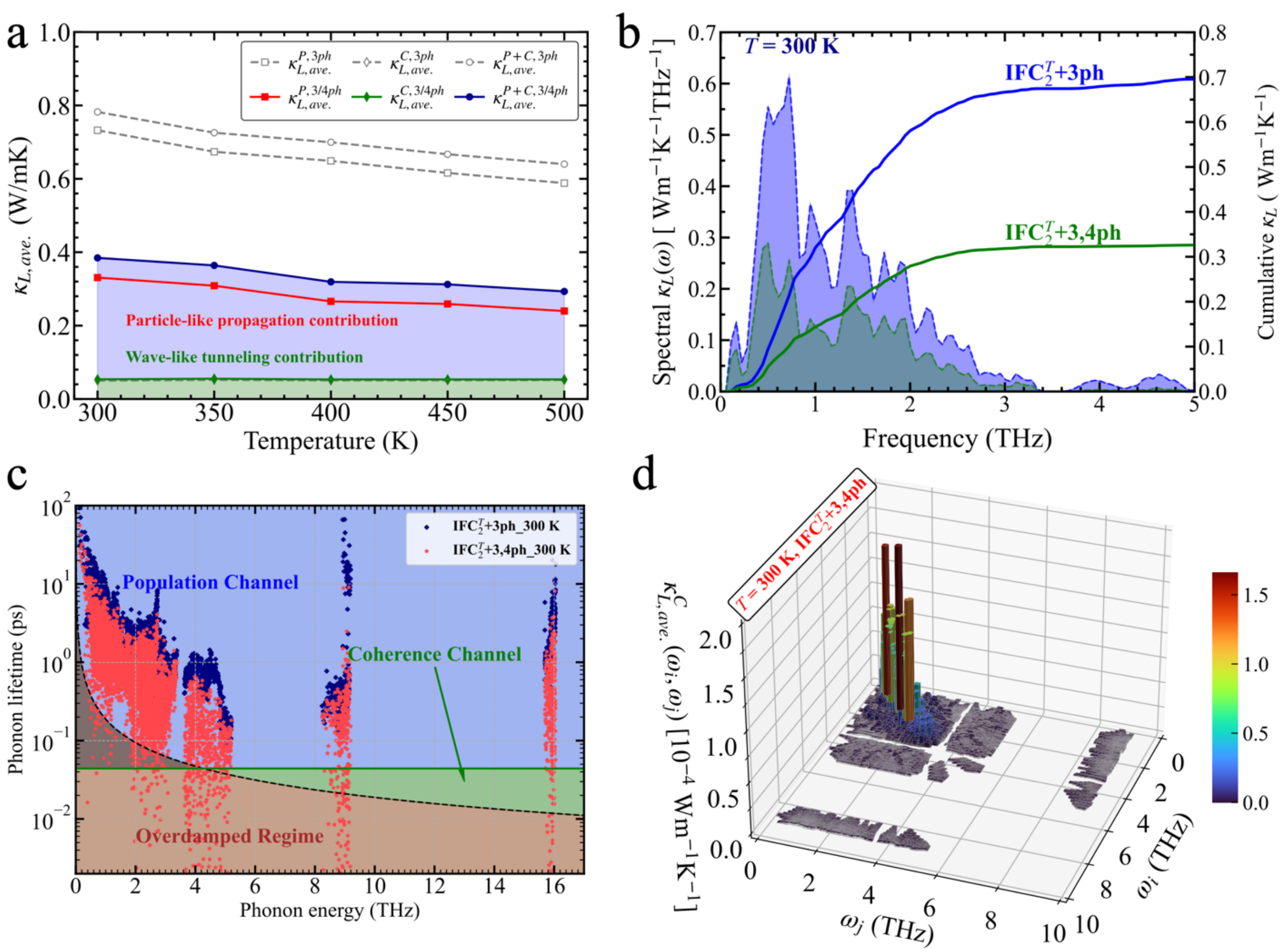

**Fig. 6. Lattice thermal conductivity and phonon scattering rates calculated using zero-K anharmonic IFCs. a** Calculated temperature-dependent averaged thermal conductivity: population and coherence contributions, total thermal conductivity, considering 3ph and/or 4ph scattering processes using zero-K IFCs. The pink shaded area indicates the particle-like phonon propagation contribution, and the light blue shaded area represents the wave-like phonon tunnelling contribution. **b** Calculated phonon lifetimes considering 3ph or 3,4ph scattering rates using zero-K anharmonic IFCs at 300 K. **c** Calculated averaged spectral and cumulative thermal conductivity from particle-like phonon propagation channel, considering only 3ph, and both 3ph and 4ph scattering processes, using zero-K

anharmonic IFCs at 300 K. **d** Calculated 2D modal coherence conductivity from wave-like phonon tunnelling channel, considering both 3ph and 4ph processes, using zero-K anharmonic IFCs at 300 K.

Given that anharmonic interatomic force constants (IFCs) may be sensitive to temperature[68], we next delve into how temperature-induced shifts in these IFCs impact the thermal transport properties of cubic $FAPbI_3$. To compare with the results based on temperature-dependent IFCs, we calculated the thermal transport properties using zero-Kelvin anharmonic interatomic force constants (IFCs), as illustrated in Figs. 6(a-d). Interestingly, while the third-order interatomic force constants (IFCs) are insensitive to temperature variations, the fourth-order IFCs demonstrate considerable sensitivity. Specifically, when including both three-phonon (3ph) and four-phonon (4ph) scatterings, the lattice $\kappa_{\mathrm{L}}^{\mathrm{P}}$ is predicted to be 0.33 $W/mK$ using zero-Kelvin anharmonic interatomic force constants (IFCs), as shown in Fig. 6(a). In contrast, employing fully temperature-dependent IFCs results in a calculated lattice $\kappa_{\mathrm{L}}^{\mathrm{P}}$ of 0.60 W/mK, as depicted in Fig. 3(a). This significant reduction in $\kappa_{\mathrm{L}}^{\mathrm{P}}$ due to the temperature effect is further illustrated in the spectral and cumulative $\kappa_{\mathrm{L}}^{\mathrm{P}}(\omega)$ plots in Fig. 6(b). Moreover, using the zero-K anharmonic IFCs leads to numerous phonons entering the overdamped regime, characterized by lifetimes less than the Ioffe-Regel limit, thereby questioning the validity of the Wigner transport equation in modelling thermal transport[38,39], as depicted in Fig. 6(c). This phenomenon underscores the critical importance of extracting fully temperature-dependent IFCs when assessing thermal transport and phonon-related properties in hybrid organic-inorganic perovskites.

Despite the significant increase in anharmonic scattering rates, the coherence contributions to the total $\kappa_{\mathrm{L}}$ remain almost unchanged [see Figs. 3(a) and 6(a) and (d)]. Again, the minor contributions from the wave-like phonon channel are attributed to the large inter-band spacings resulting from the lighter elements of $FA^{+}$ cations. The dominant coherence contributions are evidently from quasi-degenerate phonons with frequencies below 2 THz, as depicted in Fig. 6(d). Again, this observation underscores the pivotal role of $FA^+$ cations in maintaining the good crystal nature and dominant particle-like phonon propagation in cubic $FAPbI_3$. A recent study has highlighted that the displacement amplitude used in force constant extraction is sensitive to the resulting thermal conductivity[69]. To assess this effect, we further calculated the thermal conductivity using higher-order force constants obtained with different displacement amplitudes, specifically, 0.10 Å and 0.17 Å. The room-temperature thermal conductivities obtained from these setups are 0.37 W/m·K and 0.305 W/m·K, respectively. While these values are close to the thermal conductivity of 0.33 W/m·K calculated using a 0.15 Å displacement at 300 K, they remain significantly lower than the value of 0.59 W/m·K obtained using fully temperature-dependent force constants. This finding again underscores the importance of using temperature-consistent force constants when evaluating thermal transport properties in hybrid organic–inorganic perovskites.

## Experimental and theoretical Thermal conductivity

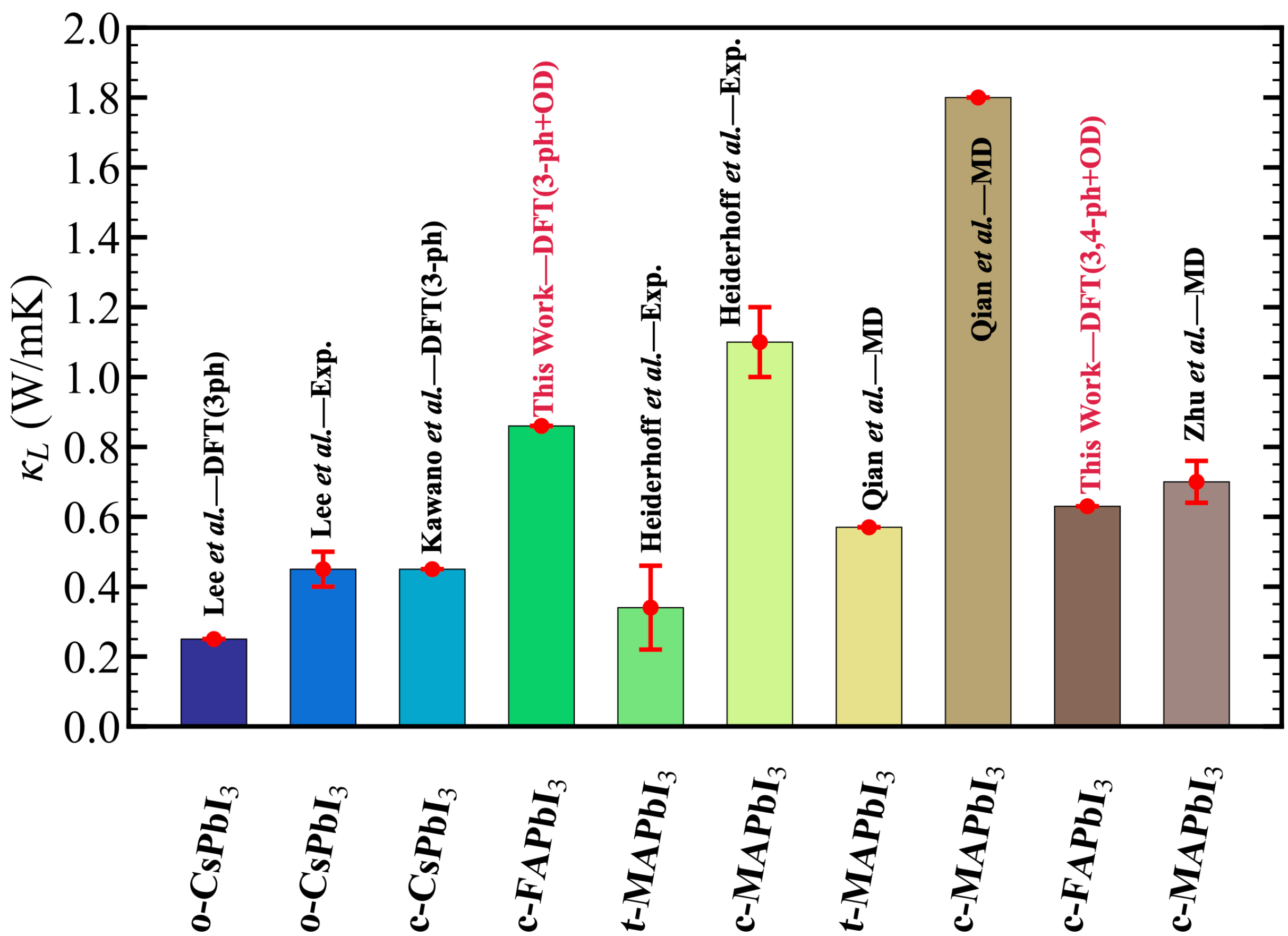

**Fig. 7. Comparison of thermal conductivity obtained from both experiments and theories.** Comparison of total lattice thermal conductivity $\kappa_L$ calculated in cubic $FAPbI_3$ with other theoretically predicted and experimentally measured ultra-low thermal conductivities in various inorganic and hybrid organic-inorganic perovskites[42,43,51,52,70]. Here, 'DFT(3ph)' denotes the value obtained by considering only three-phonon scatterings based on DFT theory, while 'DFT(3,4ph)' accounts for both three- and four-phonon scatterings. 'OD' refers to the off-diagonal terms of heat flux operators, and 'MD' stands for molecular dynamics simulation. o, t and c indicate Orthorhombic, Tetragonal and cubic phase, respectively. The red lines represent the error bars.

To confirm the accuracy of the predicted thermal conductivity $\kappa_L$ of cubic $FAPbI_3$, we compare it with experimental or theoretical $\kappa_L$ values reported for other perovskites[42,43,51,52,70] at room temperature, as illustrated in Fig. 7. We begin by comparing the thermal conductivity $\kappa_L$ of $CsPbI_3$ in various phases, both experimentally and theoretically, to that of cubic $FAPbI_3$, where only three-phonon (3ph) scatterings are considered. Our predicted room-temperature total $\kappa_L$ for cubic $FAPbI_3$ is 0.86 $Wm^{-1}K^{-1}$ when considering only 3ph scatterings, significantly higher than that reported for o/c-$CsPbI_3$. This discrepancy can be attributed to the low-frequency rattling-like modes induced by the heavy Cs atoms at A-site in $CsPbI_3$, which contribute to strong scatterings and consequently suppress thermal transport[52]. In cubic $FAPbI_3$, however, the A-site elements, i.e., $FA^+$ cations, predominantly contribute to the high-frequency modes (>3.5 THz), as depicted in Fig. 1(d). This characteristic reduces scatterings between A-site-dominated modes and primary heat carriers, i.e., low-frequency dispersive phonon modes. Subsequently, we compare the experimentally and theoretically determined $\kappa_{\mathrm{L}}$ of different phases of $MAPbI_3$ with that of cubic $FAPbI_3$, considering both three-phonon (3ph) and four-phonon (4ph) scatterings. We observe a significant increase in the thermal conductivity $\kappa_{\mathrm{L}}$ of crystalline $MAPbI_3$ during the phase transition from tetragonal to cubic, as evidenced by both molecular dynamics (MD) simulations[42] and experimental studies[70]. Given that the mass weight of $FA^+$ cations is larger than that of $MA^+$, the predicted $\kappa_{\mathrm{L}}$ of cubic $FAPbI_3$ is expected to lie between that of cubic $MAPbI_3$ and tetragonal $MAPbI_3$. Indeed, our predicted $\kappa_{\mathrm{L}}$ of cubic $FAPbI_3$, $0.63\ \mathrm{Wm^{-1}K^{-1}}$, falls between $0.34 \pm 0.12$ (tetragonal phase) and $1.1 \pm 0.1\ \mathrm{Wm^{-1}K^{-1}}$(cubic phase)[70]. Furthermore, our predicted value in cubic $FAPbI_3$ is closely aligned with the $\kappa_{\mathrm{L}}$ of cubic $MAPbI_3$ as predicted using molecular dynamics by Zhu et al.[43]. Overall, the predicted total $\kappa_L$ of cubic crystalline $FAPbI_3$ in this work is reasonably reliable and can be validated by future experimental investigations.

In summary, we have employed a first-principles-based framework that integrates temperature-dependent effective potential with the linearized Wigner transport formula, accounting for both three-phonon (3ph) and four-phonon (4ph) scatterings, to elucidate the microscopic mechanisms of thermal transport in cubic $FAPbI_3$. Our findings reveal that at 0K, cubic $FAPbI_3$ exhibits dynamical instability primarily due to the strong random orientation of $FA^+$ cations. This strong random orientation of $FA^+$ cations also trigger the tilting of $PbI_6$ octahedra, leading to the deformation of cubic $FAPbI_3$ into an unfavorable phase (δ phase) as temperature decreases.

Utilizing the Wigner transport formula, and accounting for both three-phonon (3ph) and four-phonon (4ph) scatterings, we observed an ultra-low thermal conductivity of 0.63 $\mathrm{Wm^{-1}K^{-1}}$ at room temperature for cubic $FAPbI_3$. Analysis of the COHP/IpCOHP results and the projected atomic participation ratio in cubic $FAPbI_3$ suggests that the ultra-low thermal conductivity can primarily be attributed to the $[PbI_3]^-$ units, rather than the $FA^+$ cations. This phenomenon is driven by the presence of Pb(s)-I(p) antibonding states in the $[PbI_3]^-$ units, contributing to weak bonding and strong anharmonicity.

Despite its complex structure, containing 12 atoms in the primitive cell and organic cations, and marked strong anharmonicity, the particle-like phonon propagation channel effectively explains the observed thermal conductivity in cubic $FAPbI_3$. This unique behavior can be ascribed to the presence of light elements in the organic cations, i.e., $FA^+$ cations, which lead to high-frequency phonons and significant inter-band spacings. The presence of $FA^+$ cations in cubic $FAPbI_3$ has little impact on phonon population contributions and contribute negligibly to the coherence contributions in the total $\kappa_{\mathrm{L}}$.

Furthermore, our results highlight the importance of employing fully temperature-dependent IFCs to accurately evaluate thermal transport and phonon-related properties in hybrid organic-inorganic perovskites. Finally, by comparing our predicted thermal conductivity of cubic $FAPbI_3$ with other perovskites, we confirm that our predicted total $\kappa_{\mathrm{L}}$ of cubic crystalline $FAPbI_3$ is reasonably reliable. Our work unveils the microscopic mechanisms of heat conduction physics in cubic $FAPbI_3$, paving the way to understand thermal transport in other hybrid organic-inorganic compounds.

## Methods

### First-principles calculations and Compressive Sensing Technique

All the density functional theory (DFT)[71] calculations of cubic crystalline $FAPbI_3$ were performed using the Vienna Ab initio Simulation Package (VASP)[72]. The projector-augmented wave (PAW) pseudopotentials were used to explicitly treat the valence states of C, N, H, Pb, and I atoms, considering the electron configurations ($2s^22p^2$), ($2s^22p^3$), ($1s^1$), ($5d^{10}6s^26p^2$), and ($4d^{10}5s^25p^5$) electrons, respectively. The Perdew-Burke-Ernzerhof (PBE)[73] functional within the generalized gradient approximation (GGA)[74] framework for the exchange-correlation functional was employed for all DFT calculations. Additionally, the optB86b-vdW method[75] was utilized to accurately describe the van der Waals (vdW) interactions within our computational framework. For structural optimization, a $\Gamma - center$ 10×10×10 Monkhorst-Pack *k*-mesh and a kinetic energy cutoff of 700 eV were utilized to sample the Brillouin zone in the primitive cell containing 12 atoms. The force convergence threshold was set to $10^{-5}$ eV·Å$^{-1}$ for structural optimization, and a tight energy convergence criteria of $10^{-8}$ eV was employed for both structural relaxation and static DFT calculations. The fully optimized average lattice constant ($a$ = 6.3807 Å) agrees well with the experimentally reported values ($a$ = 6.3620 Å) at room temperature for cubic crystalline $FAPbI_3$[45]. Note that in current work, we only focus on simulating the anharmonic lattice dynamics and thermal transport behavior from the experimentally observed cubic crystalline FAPbI3 with space group of $Pm\bar{3}m$. Given that the impact of thermal expansion on phonon frequency shifts in cubic crystalline $FAPbI_3$ is relatively minor compared to the effect of phonon renormalization due to anharmonicity (i.e., as captured by the TDEP method used in this work), we have opted not to include thermal expansion in the present calculations [For more details and discussion, see **Supplementary Figure 8** in the Supporting Information.].

The zero-Kelvin harmonic interatomic force constants (IFCs) were computed using the finite-displacment approach[76], implemented in ALAMODE package[77], ultilizing a $3\times3\times3$ supercell containing 324 atoms. Note that the random-displacement method combined with a least-squares fitting approach in ALAMODE[77] using 800 configurations was employed to extract the zero-K second-order force constants. As a result, the phonon dispersion may differ slightly from that obtained using the finite-displacement method in Phonopy[78]. To accurately and efficiently extract the anharmonic interatomic force constants (IFCs), particularly 3rd and 4th order terms, the Compressive Sensing Lattice Dynamics (CSLD) method[79,80] was ultilized. The CSLD method efficiently selects the physically significant terms from a large set of irreducible anharmonic interatomic force constants (IFCs), utilizing a small displacement-force dataset[27]. To extract zero-Kelvin anharmonic interatomic force constants (IFCs), we generated a set of 400 atomic structures from an equilibrium 3×3×3 supercell. Each structure was subjected to a uniform displacement of 0.15 Å for all atoms along random directions, achieved using the random number method. We would like to highlight that the average atomic displacement at 300 K is approximately 0.39 Å in cubic crystalline $FAPbI_3$. Considering the strong anharmonicity and the presence of flat double-well potential energy surfaces in this material, a uniform displacement of 0.15 Å was selected to enhance the signal-to-noise ratio and reduce fitting errors in the extraction of higher-order force constants at zero K. However, for the finite-temperature anharmonic interatomic force constants (IFCs), we generated a set of 400 atomic structures with the atoms displaced according to a harmonic canonical ensemble at finite temperatures[81,82]. Subsequently, the obatined 400 atomic structures were used to generate displacement-force dataset through precise DFT calculations with a $\Gamma - center$ 2×2×2 Monkhorst-Pack $k$-point density grid. Finally, the harmonic interatomic force constants (IFCs), either at zero-Kelvin or finite temperatures, along with the displacement-force dataset, were utilized to extract anharmonic IFCs up to the sixth order. The anharmonic IFCs were extracted using the least absolute shrinkage and selection operator (LASSO) technique[83], applying real-space cutoff radii of 6.88 Å , 5.82 Å, 3.17 Å and 3.17 Å for cubic, quartic, quintic and sextic IFCs, respectively. Here, we would like to emphasize that the cutoff distances of the third- and fourth-order IFCs were confirmed through convergence testing to yield negligible changes in thermal conductivity beyond these values [see **Supplementary Figure 9** in Supplemental Information]. In this work, the IFCs fitting process was conducted using the ALAMODE package[27,77].

**Temperature-dependent Effective Potential Method**

To account for the temperature effect on phonon normal modes, we utilize the temperature-dependent effective potential (TDEP)[28,29] approach to fit first-principles forces to an effective Hamiltonian ($H$),

$$H = U_0 + \sum_i \frac{\boldsymbol{P}_i^2}{2m_i} + \frac{1}{2}\sum_{ij\alpha\beta} \Phi_{ij}^{\alpha\beta} u_i^\alpha u_j^\beta, \quad (1)$$

where $U_0$ is the potential energy of static lattice, $\boldsymbol{p}_i$, $m_i$ and $u_i$ are the momentum, atomic mass and diaplacment associated with atom $i$, respectively. $\Phi_{ij}^{\alpha\beta}$ is the effective harmonic IFCs, i.e., second-order force constant, associated with the Cartesian indices.

To obtain the displacment-force dataset, we computed precise DFT forces on atoms within perturbed supercells, which were generated using stochastic sampling of a canonical ensemble[81,82]. The Cartesian displacement ($u_i^\alpha$) is normally distributed around the mean thermal dispacement and can be expressed as:

$$u_i^\alpha = \sum_q e_q^{i\alpha} \langle A_{iq} \rangle \sqrt{-2ln\zeta_1}\,\sin(2\pi\zeta_2), \quad (2)$$

with the thermal amplitude $\langle A_{iq} \rangle$ given by[81,82,84]:

$$\langle A_{iq} \rangle = \sqrt{\frac{\hbar(2n_q^0+1)}{2m_i\omega_q}}, \tag{3}$$

where the phonon mode $q$ is a composite index of the wavevector $\boldsymbol{q}$ and phonon branch $s$, $e_q$ is phonon eigenvector, $\zeta_1$ and $\zeta_2$ are stochastically sampled numbers between 0 and 1, $\hbar$ is the Planck constants, $n_q$ is the occupation number following Bose-Einstein statistics, and $\omega_q$ is the phonon frequency.

In this study, we utilize a 3×3×3 supercell of cubic $FAPbI_3$ and perform calculation iteratively, starting from 600 thermally perturbed snapshots. At each temperature (300, 350, 400, 450 and 500 K), the last iteration is conducted using 3,600 snapshots to ensure the convergence of finite-temperature IFCs. Each iteration involves key procedures such as computing phonon normal modes, genrating perturbed snapshots, calculating precise DFT forces, and fitting effective IFCs. It is worth noting that the force constants were extracted from thermally sampled configurations generated by stochastic sampling of a canonical ensemble, thereby inherently accounting for both the thermal motion and rotational degrees of freedom of the $FA^+$ cations. In this study, the temperature-dependent effective potential calculations were performed using both the TDEP[28,29] and ALAMODE package[77]. It's worth noting that in this study, we utilized our in-house code[30,36] for the transformation of force constants between the ALAMODE[27,77] and ShengBTE[85] packages.

**Intrinsic and Extrinsic Phonon Scattering Rates**

Using Fermi's golden rule within time-dependent perturbation theory[61], the intrinsic scattering rates for three- (3ph) $\Gamma_q^{3ph}$ and four-phonon (4ph) $\Gamma_q^{4ph}$ processes are determined by treating the cubic and quartic anharmonic terms as perturbations. Under the single-mode relaxation time approximation (SMRTA) treatment, the intrinsic scattering rates $\Gamma_q^{3ph}$ and $\Gamma_q^{4ph}$ can be formulated as[54,57,61]

[35]

$$\Gamma_q^{3\mathrm{ph}} = \sum_{q'q''} \left\{ \frac{1}{2}\left(1 + n_{q'}^0 + n_{q''}^0\right)\zeta_- + \left(n_{q'}^0 - n_{q''}^0\right)\zeta_+ \right\}, \tag{4}$$

$$\Gamma_q^{4\mathrm{ph}} = \sum_{q'q''q'''} \left\{ \frac{1}{6}\frac{n_{q'}^0 n_{q''}^0 n_{q'''}^0}{n_q^0}\zeta_{--} + \frac{1}{2}\frac{\left(1+n_{q'}^0\right)n_{q''}^0 n_{q'''}^0}{n_q^0}\zeta_{+-} + \frac{1}{2}\frac{\left(1+n_{q'}^0\right)\left(1+n_{q''}^0\right)n_{q'''}^0}{n_q^0}\zeta_{++} \right\}, \tag{5}$$

with

$$\zeta_\pm = \frac{\pi\hbar}{4N}\left|V^{(3)}\left(q, \pm q', -q''\right)\right|^2 \Delta_\pm \frac{\delta\left(\Omega_q \pm \Omega_{q'} - \Omega_{q''}\right)}{\Omega_q \Omega_{q'} \Omega_{q''}}, \tag{6}$$

and

$$\zeta_{\pm\pm} = \frac{\pi\hbar^2}{8N^2}\left|V^{(4)}\left(q, \pm q', \pm q'', -q'''\right)\right|^2 \Delta_{\pm\pm} \frac{\delta\left(\Omega_q \pm \Omega_{q'} \pm \Omega_{q''} - \Omega_{q'''}\right)}{\Omega_q \Omega_{q'} \Omega_{q''} \Omega_{q'''}}, \tag{7}$$

where $\Omega_q$ is the finite-temperature harmonic phonon frequency, $V^{(3)}\left(q, \pm q', -q''\right)$ and $V^{(4)}\left(q, \pm q', \pm q'', -q'''\right)$ are the reciprocal representation of 3rd- and 4th-order IFCs, respectively[86], for both 3ph and 4ph scattering processes, energy and momentum conservation are enforced by delta function $\delta$ and Kronecker delta $\Delta$, respectively.

The extrinsic phonon scattering arising from naturally occurring isotopes, indicated as $\Gamma_q^{isotope}$, can be formulated as[87]

$$\Gamma_q^{\mathrm{isotope}} = \frac{\pi\Omega_q^2}{2N} \sum_{i \in u.c.} g(i)\left|e_q^*(i) \cdot e_{q'}(i)\right|^2 \delta\left(\Omega - \Omega'\right), \tag{8}$$

where $g(i)$ is the Pearson deviation coefficient[87]. Using Matthiessen's rule, the total phonon scattering rate $\Gamma_q$ for a phonon mode $q$ can be formulated as

$$\Gamma_q = \Gamma_q^{3\mathrm{ph}} + \Gamma_q^{4\mathrm{ph}} + \Gamma_q^{\mathrm{isotope}}, \tag{9}$$

**Linearized Wigner Transport Formula**

To accurately evaluate thermal transport in cubic $FAPbI_3$, we utilize the linearized Wigner transport equation[38,39] to consider both the contributions from particle-like phonon propagation $\kappa_{\mathrm{L}}^{\mathrm{P}}$ and wave-like tunnelling of phonons $\kappa_{\mathrm{L}}^{\mathrm{C}}$ to total thermal conductivity $\kappa_{\mathrm{L}}$. The Wigner transport equation has been widely applied to reproduce and explain experimentally observed thermal conductivity in materials ranging from simple systems[30,56] to complex systems[38,39,62]. Its validity has also been confirmed through comparison with Green–Kubo calculations, which account for all contributions to heat transport, including cases with non-Lorentzian phonon spectral functions[88]. Under the SMRTA treatment, the linearized Wigner transport equation can be formulated as[38,39]

$$\begin{aligned}\kappa_{\mathrm{L}}^{\mathrm{P/C}} = {} & \frac{\hbar^2}{k_B T^2 V N} \sum_{\boldsymbol{q}} \sum_{j,j'} \frac{\Omega_{\boldsymbol{q}j} + \Omega_{\boldsymbol{q}j'}}{2} \boldsymbol{\upsilon}_{\boldsymbol{q}jj'} \otimes \boldsymbol{\upsilon}_{\boldsymbol{q}j'j} \\ & \bullet \frac{\Omega_{\boldsymbol{q}j} n_{\boldsymbol{q}j}(n_{\boldsymbol{q}j}+1) + \Omega_{\boldsymbol{q}j'} n_{\boldsymbol{q}j'}(n_{\boldsymbol{q}j'}+1)}{4(\Omega_{\boldsymbol{q}j} - \Omega_{\boldsymbol{q}j'})^2 + (\Gamma_{\boldsymbol{q}j} + \Gamma_{\boldsymbol{q}j'})^2} (\Gamma_{\boldsymbol{q}j} + \Gamma_{\boldsymbol{q}j'}),\end{aligned} \tag{10}$$

where $k_B$ is the Boltzmann constant, $V$ is the volume of primitive cell, $T$ is the absolute temperature, $N$ is the number of sampled phonon wave vectors and $\boldsymbol{\upsilon}$ is the group velocity matix, including both diagonal and off-diagonal terms[89]. When $j = j'$, it corresponds to diagonal terms of heat flux operators, contributing to populations' contribution, $(\kappa_{\mathrm{L}}^{\mathrm{P}})$. Otherwise, it corresponds to off-diagonal terms of heat flux operators, providing the coherences' contribution $(\kappa_{\mathrm{L}}^{\mathrm{C}})$ in Eq. (10). To solve Eq. (10), we ultilize a $\boldsymbol{q}$ mesh of 12×12×12 for the both 3ph and 4ph scattering processes, with scalebroad parameters set at 0.1. The **q**-mesh and scalebroad parameter used in current work for the thermal conductivity of cubic $FAPbI_3$ was verified through convergence testing to yield results within 2% and 2% of denser grids and larger scalebroad parameter [see **Supplementary Figure (10-11)** in Supplemental Information]. Note that we adopt an iterative scheme to address the diagonal terms of heat flux operators in three-phonon (3ph) scattering processes. In contrast, the SMRTA treatment is employed to handle the four-phonon (4ph) scattering processes, considering the extremely high computer memory demands[86]. In this work, thermal conductivity calculations, including populations' and coherences' contributions, were performed using the ShengBTE[85] and FourPhonon[86,90] packages, along with our in-house code[30,36].

## DATA AVAILABILITY

Data that support the findings of this study will be available from the corresponding authors upon reasonable request.

## CODE AVAILABILITY

The open-source codes can be found as following: Alamode is available at https://github.com/ttadano/alamode, ShengBTE is available at https://www.shengbte.org and FOURPHONON is available at https://github.com/FourPhonon/FourPhonon. The in-house codes will be available from the corresponding authors upon reasonable request.

## ACKNOWLEDGEMENT

G.H. and J.Z. acknowledge funding by the U.S. Department of Energy, Office of Science, Office of Basic Energy Sciences, Materials Sciences and Engineering Division, under Contract No. DE-AC02-05-CH11231: Materials Project program KC23MP. R.G. acknowledges support from the Excellent Young Scientists Fund (Overseas) of Shandong Province (2022HWYQ091) and the Initiative Research Fund of Shandong Institute of Advanced Technology (2020107R03). B.H. acknowledges the financial support from the Science and Technology Planning Project of Guangdong Province, China (Grant No. 2017A050506053), the Science and Technology Program of Guangzhou (No. 201704030107), and the Hong Kong General Research Fund (Grants No. 16214217 and No. 16206020). This paper was also supported in part by the Project of Hetao Shenzhen-Hong Kong Science and Technology Innovation Cooperation Zone (HZQB-KCZYB2020083). C.L. acknowledges the support from the Sinergia project of the Swiss National Science Foundation (grant number CRSII5_189924). Z.C. acknowledges support from the Fundamental Research Fund for Zhejiang Ocean University (Grant No. JX6311181423).

## AUTHOR CONTRIBUTIONS

J.Z. contributed to the methodology, software, all calculations, analysis and writing of the original draft. Z.C., ChangpengL, ChongjiaL, and Y.Z. contributed to analysis and writing of the manuscript. B.H., R.G. and G.H. were responsible for supervision, writing and analysis throughout. All authors reviewed the final manuscript.

## Competing interests

The Authors declare no Competing Interests.